\documentclass[universe,submit,article,moreauthors]{Definitions/mdpi} 
\firstpage{1} 
\makeatletter
\let\linenumbers\relax
\makeatother
\pubvolume{1}
\issuenum{1}
\articlenumber{0}
\pubyear{2026}
\copyrightyear{2026}
\datereceived{ } 
\daterevised{ } % Comment out if no revised date
\dateaccepted{ } 
\datepublished{ } 

\Title{From Observation to Ground Truth: Reconstruction Losses of Galaxy Imaging Under Heteroscedastic Noise}

\Author{Renhao Ye $^{1,2}$\orcidA{}, Shiyin Shen $^{1,3}$\orcidB{} and Quanfeng Xu $^{1,2}$\orcidC{}}

\AuthorNames{Renhao Ye, Shiyin Shen and Quanfeng Xu}

\address{%
$^{1}$ \quad Shanghai Astronomical Observatory, Chinese Academy of Sciences, 80 Nandan Road, Shanghai 200030, China\\
$^{2}$ \quad School of Astronomy and Space Science, University of Chinese Academy of Sciences, Beijing 100049, China\\
$^{3}$ \quad Shanghai Key Laboratory for Astrophysics, Shanghai Normal University, Shanghai 200234, China}

\corres{Correspondence: ssy@shao.ac.cn}

\abstract{Deep learning is widely used to analyze galaxy images, and the reconstruction loss determines which image features a model prioritizes during training. Mean squared error (MSE) and mean absolute error (MAE) correspond to homoscedastic Gaussian and Laplace likelihoods, respectively, whereas the inverse-variance-weighted $\chi^2$ loss accounts for the spatially varying uncertainties of heteroscedastic astronomical noise. Comparing how closely these objectives recover the underlying signal requires a ground truth (GT), which real observations cannot provide. We therefore generate noise-free galaxy images with IllustrisTNG, SKIRT, and \textsc{GalaxyGenius}, convolve them with a point-spread function (PSF) to define the GT, and construct noisy simulated observations. We train otherwise identical variational autoencoders (VAEs) with the three reconstruction losses and evaluate their outputs against the GT. Over most of the evaluated signal-to-noise ratio (SNR) range, the $\chi^2$-trained model yields lower relative reconstruction errors than the MSE- and MAE-trained models, indicating that inverse-variance weighting improves galaxy-image reconstruction under the heteroscedastic noise considered here.}


\begin{document}
\nolinenumbers
%%%%%%%%%%%%%%%%%%%%%%%%%%%%%%%%%%%%%%%%%%
\section{Introduction}
Galaxy images encode information about galaxy formation and evolution through their brightness distributions and multiscale spatial structures. Deep learning has been applied to these images for denoising\cite{liu_astronomical_2025a,mirzoyan_enhancing_2025a}, super-resolution\cite{koopmans_superresolving_2025}, deconvolution\cite{sukurdeep_astroclearnet_2025,sukurdeep_imagemm_2025a}, and representation learning\cite{mohale_enabling_2024,himes_multimodal_2025,howie_deciphering_2025}. In such applications, the reconstruction loss determines which image properties the model prioritizes during training. Galaxy images span a wide surface-brightness range, from bright central cores to faint outer disks, rings, and tidal features\cite{golini_lights_2025,mosenkov_unveiling_2022,li_cosmic_2025,martinez-delgado_stellar_2025b}. These faint structures have low signal-to-noise ratios (SNRs) and are often obscured by noise, making their recovery sensitive to the choice of reconstruction loss.

Fundamentally, a reconstruction loss corresponds to an assumption about the pixel-noise distribution. Mean squared error (MSE) and mean absolute error (MAE) correspond to homoscedastic Gaussian and Laplace likelihoods, respectively; both effectively assume a spatially constant noise scale\cite{zhao_loss_2018}. This simplification was common in early deep image-reconstruction work on standard natural-image datasets\cite{dong_image_2016}. The advantage of the inverse-variance-weighted $\chi^2$ loss investigated here does not arise from the squared-error form itself, but from incorporating spatially varying uncertainties into the likelihood weighting. Therefore, our comparison should be interpreted as a test of uncertainty-aware weighting under heteroscedastic noise rather than as evidence that squared error intrinsically outperforms conventional reconstruction objectives.

In low-photon-count imaging, however, photon counts follow Poisson statistics, and their variance grows with the signal; the resulting noise is therefore heteroscedastic\cite{howell_handbook_2006}. In imaging domains including fluorescence microscopy, likelihood-based approaches account for an estimated per-pixel noise standard deviation ($\sigma$) and can improve denoising relative to homoscedastic MSE\cite{kendall_what_2017b,khademi_selfsupervised_2021}. Astronomical data-reduction pipelines likewise provide per-pixel uncertainty products, often as inverse-variance maps, that characterize instrumental noise, sky-background uncertainty, read noise, and related terms\cite{lang_legacypipe_2025,dey_overview_2019a}. Converting these products to $\sigma$ makes the $\chi^2$ loss a natural reconstruction objective. Nevertheless, $\chi^2$ has seen limited use in astronomical image restoration\cite{sukurdeep_astroclearnet_2025,sukurdeep_imagemm_2025a}, and it has not been directly compared with MSE and MAE.

A real observation is a single noisy realization of its underlying signal. Because real observations do not provide a ground truth (GT), they cannot directly determine which of the three losses yields reconstructions closest to the noise-free signal. We therefore use cosmological simulations and radiative-transfer tools\cite{nelson_illustristng_2021,nelson_first_2019,camps_skirt_2020,zhou_galaxygenius_2025} to generate noise-free galaxy images with realistic morphologies. From each GT image, we simulate a noisy observation and its per-pixel uncertainty map. Under a fixed network architecture and training setup, we vary the reconstruction loss among MSE, MAE, and $\chi^2$ and compare the resulting reconstructions against the GT.

\section{Methodology}
\label{sec:method}
\begin{figure}[t]
\centering
\includegraphics[width=\textwidth]{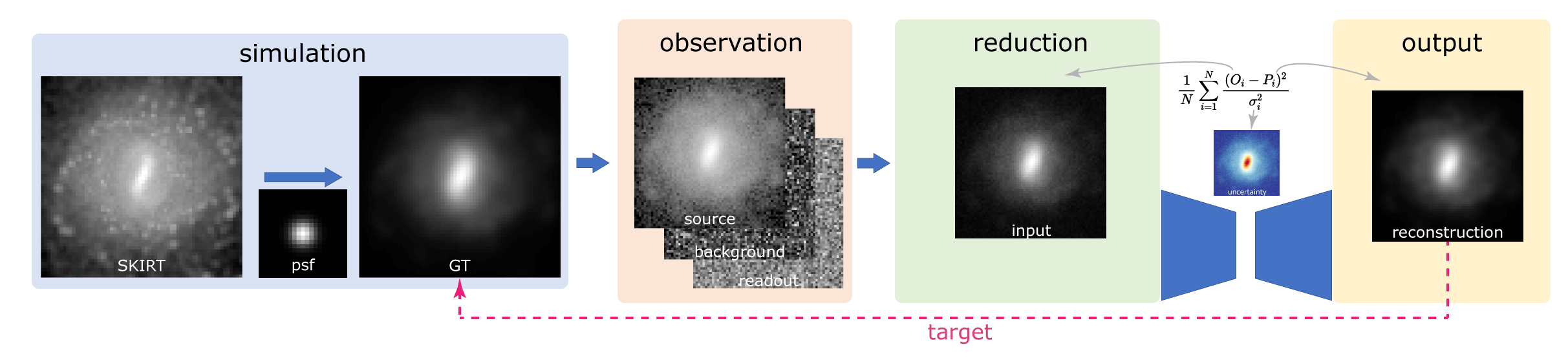}
\caption{Overview of the pipeline. \textit{Simulation}: a SKIRT galaxy image is convolved with the point-spread function (PSF) to form the GT. \textit{Observation}: the forward model draws source, background, and readout noise. \textit{Reduction}: this process yields the background-subtracted observation and its per-pixel uncertainty map. \textit{Output}: the model reconstructs the observation. The magenta dashed line denotes the final evaluation against the GT.}
\label{fig:pipeline}
\end{figure}

\subsection{SKIRT GT Simulation}
\label{sec:simulation}
We generate the GT images by combining cosmological hydrodynamical simulations with radiative transfer. IllustrisTNG is a suite of cosmological magnetohydrodynamical simulations that self-consistently follows the formation and evolution of galaxies; its high-resolution TNG50 run supplies the resolved galaxy morphologies used here\cite{nelson_illustristng_2021,nelson_first_2019,pillepich_first_2019}. SKIRT is a Monte Carlo radiative-transfer code that propagates stellar emission through dust absorption and scattering to synthesize images in specified bands\cite{camps_skirt_2020}. \textsc{GalaxyGenius} connects the hydrodynamical simulation to SKIRT and produces mock images for a specified telescope\cite{zhou_galaxygenius_2025}. We select 1{,}153 subhalos in snapshot 98 of IllustrisTNG50-1 with stellar masses of $10^{8}$--$10^{12}\,M_\odot$ (i.e., all subhalos within this mass range, so that the sample follows the simulation's own stellar mass function rather than a uniform sampling in $\log M$) and render each from seven random viewing angles in the $r$ band of the China Space Station Telescope (CSST)\cite{collaboration_introduction_2025}, using $10^{8}$ photon packets per SKIRT run. Although the galaxies are drawn at the redshift z=0.01 of snapshot 98, we cosmologically rescale their apparent brightness and angular size to $z=0.04$ to accelerate training and reduce computational requirements. The forward model in Section~\ref{sec:noise_model} then produces the GT and observation (Figure~\ref{fig:pipeline}).

\subsection{Heteroscedastic Observation Model}
\label{sec:noise_model}
We first convolve the noise-free source image with the instrumental PSF, modeled as a Moffat profile with $\beta=3.5$ and a full width at half maximum of $1.5''$, to obtain the GT. We add a weak first-order sky plane with a surface brightness of $22.0~\mathrm{mag\,arcsec^{-2}}$, tilted in a random direction with a peak-to-center amplitude equal to $0.2$ percent of the sky level. Let $S_i$ and $B_i$ denote the PSF-convolved source and sky electron rates at pixel $i$. For each of $n=3$ independent exposures of duration $t=60~\mathrm{s}$, we draw
\begin{equation}
Q_{ij}\sim\operatorname{Poisson}\!\left[t(S_i+B_i)\right],
\qquad
\epsilon_{ij}\sim\mathcal{N}\!\left(0,\sigma_{\mathrm{RN}}^2\right),
\end{equation}
and set $C_{ij}=Q_{ij}+\epsilon_{ij}$, where the Poisson and Gaussian random variables are independent, $j$ indexes the exposure, $\sigma_{\mathrm{RN}}=3~\mathrm{e^-}$ is the read-noise standard deviation, and the gain is unity. The background for each exposure is estimated from a companion source-free detector region that shares the same underlying tilted sky plane but has an independent realization of Poisson and read noise. We apply \texttt{Background2D} with meshes containing $8\times8$ pixels, a $3\times3$ median filter, and $3\sigma$ clipping, using the median within each mesh. The estimated background $\widehat{B}_{ij}$ is subtracted from the corresponding science exposure.

The three background-subtracted exposures are averaged and converted linearly from electron rate to nanomaggies (nMgy), a linear flux unit in the AB photometric system. The nanomaggies scale is defined such that a flux $f_{\rm nMgy}$ corresponds to an AB magnitude of $m_{\rm AB}=22.5-2.5\log_{10}(f_{\rm nMgy})$. With the photometric zero point $\mathrm{ZP}=24.6$, we define the conversion factor
\begin{equation}
k=10^{0.4(22.5-\mathrm{ZP})},
\end{equation}
in units of nanomaggies per $\mathrm{e^-\,s^{-1}}$. The delivered observation and GT are then
\begin{equation}
O_i=\frac{k}{nt}\sum_{j=1}^{n}\left(C_{ij}-\widehat{B}_{ij}\right),
\qquad
\mathrm{GT}_i=kS_i.
\end{equation}

The source shot-noise contribution to the uncertainty is estimated from the observation rather than from the GT. Let $\widehat{S}_i$ denote the nonnegative source-rate estimate obtained from $O_i$ for this purpose. After propagation through the reduction procedure, the per-pixel observation is approximated by a Gaussian centered on the GT,
\begin{equation}
O_i \sim \mathcal{N}\!\left(\mathrm{GT}_i,\sigma_i^2\right),
\end{equation}
whose variance is the sum of the source shot-noise, sky shot-noise, read-noise, and background-estimation variances. This Gaussian model approximates the underlying Poisson--Gaussian statistics and is least accurate in the lowest-count regime. For one representative exposure, with the exposure index $j$ suppressed, the background-estimation variance is approximated from the background root mean square (RMS) as
\begin{equation}
\operatorname{Var}\!\left(\widehat{B}_i\right)
\simeq \frac{\pi}{2}\frac{\mathrm{RMS}_{\mathrm{bkg},i}^2}{N_{\mathrm{pix}}},
\qquad N_{\mathrm{pix}}=8^2.
\end{equation}
The variance of the mean of the $n$ exposures, expressed in the delivered units, is therefore
\begin{equation}
\sigma_i^2
=\frac{k^2}{nt^2}
\left[t\widehat{S}_i+tB_i+\sigma_{\mathrm{RN}}^2+
\operatorname{Var}\!\left(\widehat{B}_i\right)\right]
\equiv \sigma_{\mathrm{src},i}^2+\sigma_{\mathrm{sky},i}^2
+\sigma_{\mathrm{RN},i}^2+\sigma_{\mathrm{bkg},i}^2.
\label{eq:variance_decomposition}
\end{equation}
Only the total $\sigma_i$ is retained in the per-pixel uncertainty map. Because the source term depends on the observed flux and the sky and background terms vary spatially, the noise is heteroscedastic.

\subsection{Controlled Reconstruction Objectives}
The variational autoencoder (VAE)\cite{kingma_autoencoding_2014} reconstructs its input: the single-channel noisy observation $O$ is both the network input and the reconstruction target. Its encoder, with parameters $\phi$, defines a diagonal-Gaussian posterior $q_\phi(z\mid O)$, from which the latent variable $z$ is sampled; the decoder then produces the reconstruction $P$, with $P_i$ denoting pixel $i$. All models share the same architecture and optimization settings and differ only in the reconstruction term. Under the Gaussian approximation and treating the supplied $\sigma_i$ as fixed, the per-pixel negative log-likelihood is
\begin{equation}
-\log p(O_i\mid P_i,\sigma_i) = \frac{(P_i-O_i)^2}{2\sigma_i^2} + \frac{1}{2}\log\!\left(2\pi\sigma_i^2\right).
\label{eq:nll}
\end{equation}
Here $\sigma_i$ is the propagated uncertainty at pixel $i$. It is fixed with respect to the model parameters, so the second term is constant during optimization. Dropping this additive term, multiplying by 2, and averaging over all $N$ pixels gives the inverse-variance-weighted $\chi^2$ reconstruction loss
\begin{equation}
\mathcal{L}_{\chi^2} = \frac{1}{N}\sum_{i=1}^{N}\frac{(P_i-O_i)^2}{\sigma_i^2}.
\label{eq:chi2_loss}
\end{equation}
Assuming instead a spatially uniform Gaussian noise scale, i.e., a homoscedastic $\sigma$, reduces Equation~\eqref{eq:chi2_loss} to the mean squared error up to a constant factor,
\begin{equation}
\mathcal{L}_{\mathrm{MSE}} = \frac{1}{N}\sum_{i=1}^{N}(P_i-O_i)^2,
\end{equation}
whereas a Laplace likelihood with a spatially uniform scale yields the mean absolute error,
\begin{equation}
\mathcal{L}_{\mathrm{MAE}} = \frac{1}{N}\sum_{i=1}^{N}|P_i-O_i|.
\end{equation}
Thus, the $\chi^2$ reconstruction loss retains the per-pixel measurement variance and down-weights highly uncertain pixels, whereas the MSE and MAE objectives use spatially constant noise scales and do not continuously weight pixels by their individual uncertainties.

For each choice $\mathcal{L}_{\mathrm{rec}}\in\{\mathcal{L}_{\mathrm{MSE}},\mathcal{L}_{\mathrm{MAE}},\mathcal{L}_{\chi^2}\}$, the total VAE objective is
\begin{equation}
\mathcal{L}_{\mathrm{total}}
=\mathcal{L}_{\mathrm{rec}}
+10^{-4}D_{\mathrm{KL}}\!\left[q_\phi(z\mid O)\,\|\,\mathcal{N}(0,I)\right].
\end{equation}
Here $D_{\mathrm{KL}}$ denotes the Kullback--Leibler (KL) divergence, and $I$ is the identity matrix. The KL divergence term regularizes the latent distribution by encouraging the encoder posterior $q_\phi(z|O)$to remain close to the standard normal prior, which avoids an excessively unconstrained latent space. We adopt a fixed KL weight of $10^{-4}$ for the primary comparison. Because the three reconstruction objectives have different numerical scales, however, an identical KL coefficient does not guarantee the same effective balance between reconstruction and latent regularization. We quantify this balance and test its influence in Appendix~\ref{app:kl_sensitivity}. Only the $\chi^2$ objective uses the uncertainty map for continuous inverse-variance weighting. The reconstruction model is held identical across the three objectives so that the observed differences isolate the effect of the reconstruction loss rather than of the architecture. We note, however, that the effective regularization strength and the checkpoint-selection criterion are not fully matched across the three objectives.
\section{Experiments}
\label{sec:experiments}
\subsection{Dataset}
Applying the forward model of Section~\ref{sec:noise_model} to the 8,071 simulated galaxy views yields an observation, a propagated per-pixel uncertainty map, and the GT for each view. Native images are zero-padded to at least $500$ pixels along each spatial dimension. A fractional, PSF-aware \texttt{COVERAGE} map records which output pixels are supported by native image data. This map is used only to exclude padding-affected pixels from the final evaluation. Four random $64\times64$ crops are drawn from each simulated observation without requiring a detectable ground-truth signal. The final dataset contains 8,071 observations and 32,284 crops.

The split is performed by IllustrisTNG subhalo rather than by crop, assigning 70, 15, and 15 percent of the subhalos to training, validation, and testing, respectively. This yields 807 training, 173 validation, and 173 test subhalos, corresponding to 5{,}649, 1{,}211, and 1{,}211 galaxy views and 22{,}596, 4{,}844, and 4{,}844 crops. Consequently, different views or crops of the same underlying galaxy cannot cross data partitions. The GT does not participate in training or validation; the PSF-convolved GT and \texttt{COVERAGE} map are stored only for the test set and used in the final evaluation.

\subsection{Training and Checkpoint Selection}
We train otherwise identical models with the MSE, MAE, and $\chi^2$ reconstruction losses for 60,000 optimization steps at a fixed learning rate of $10^{-4}$; the remaining optimization details are given in Appendix~\ref{app:implementation}. Checkpoints were selected by the minimum of each model's own validation reconstruction metric; the selected checkpoints occur at epochs 151, 136, and 149 for MAE, MSE, and $\chi^2$, respectively, all within the final ~20\% of training, indicating no pathological early stopping.

\begin{figure}[t]
\centering
\includegraphics[width=0.82\textwidth]{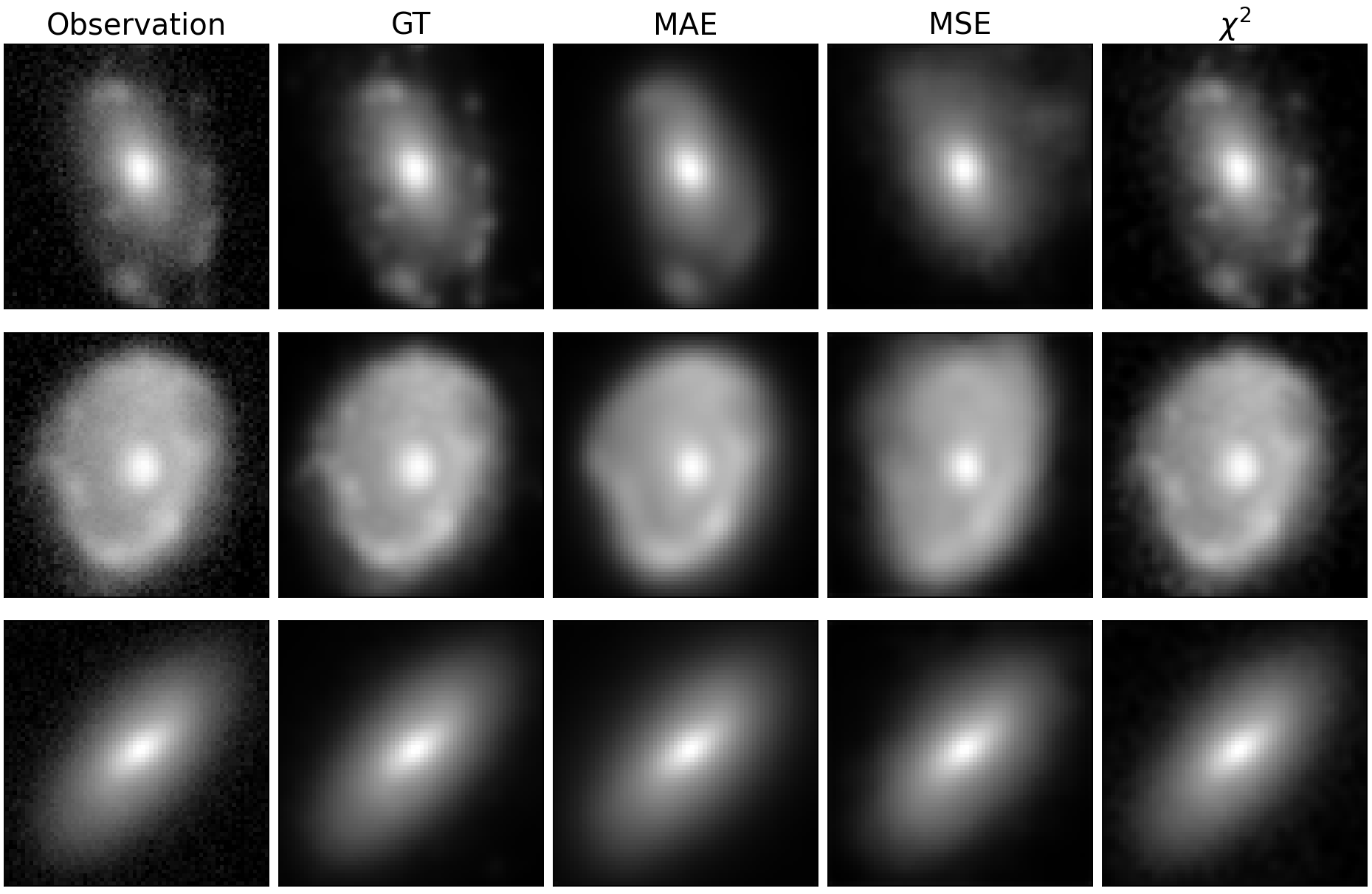}
\caption{Three randomly selected test galaxies, each centered on the galaxy. Columns show the noisy observation, GT, and reconstructions produced by models trained with the MAE, MSE, and $\chi^2$ losses.} The same color scale is used within each row. For display, the transformation $\operatorname{arcsinh}(50I)/\sqrt{50}$ is applied to each image $I$.
\label{fig:random_reconstructions}
\end{figure}
\section{Results}
\label{sec:results}
\subsection{Qualitative Galaxy-Signal Recovery}
Figure~\ref{fig:random_reconstructions} compares the noisy observation, the GT, and the three reconstructions for randomly selected test galaxies. All three models recover the dominant high-surface-brightness cores and broad galaxy structures. The MAE- and MSE-trained models produce smoother reconstructions, whereas the $\chi^2$-trained model retains more low-contrast spatial variation. Specifically, the $\chi^2$ reconstruction better preserves the string of faint clumps in the first row, the faint outer end and internal texture of the ring-like structure in the second row, and the distribution and contrast of the clumps above the broad curved structure in the third row.

\subsection{Relative Reconstruction Error across SNR}
\label{sec:evaluation}
We quantify reconstruction accuracy against the GT for pixels whose signal is at least equal to the estimated uncertainty, i.e., $\mathrm{SNR}_i\ge1$. The per-pixel SNR and relative error are
\begin{align}
\mathrm{SNR}_i = \frac{\mathrm{GT}_i}{\sigma_i}, \qquad
\varepsilon_i = \frac{|P_i-\mathrm{GT}_i|}{\max(|\mathrm{GT}_i|,\sigma_i)}.
\label{eq:relative_error}
\end{align}
The denominator is floored at the per-pixel noise $\sigma_i$ rather than clipped at any fixed value. Since only pixels with $\sigma_i>0$ and $\mathrm{SNR}_i\ge1$ are retained, for which $\mathrm{GT}_i\ge\sigma_i$, this floor equals $\mathrm{GT}_i$ on all plotted pixels and serves solely to prevent division by values below the noise level. We also exclude pixels with $\texttt{COVERAGE}<0.999$ to prevent zero-padding from affecting the metric. The eligible pixels are grouped into 30 logarithmically spaced SNR bins spanning $1\le\mathrm{SNR}\le10^{2.5}$. For each bin, we report the median and the 16th--84th percentiles of the pixel-level relative-error distribution; the median is used because the relative-error distribution is heavy-tailed at low SNR, where the mean would be dominated by a small fraction of pixels with very large errors.

\begin{figure}[t]
\centering
\includegraphics[width=0.9\linewidth]{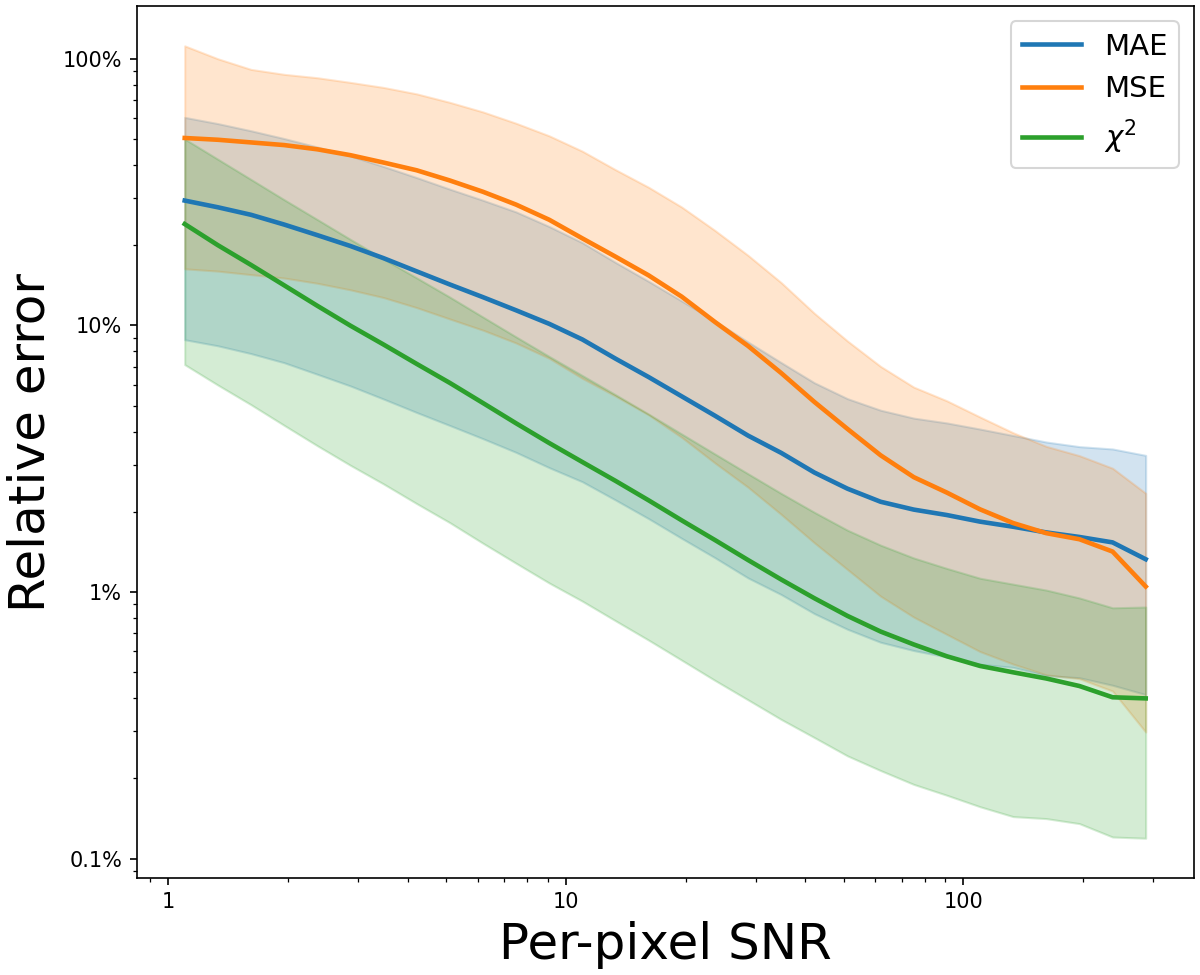}
\caption{Relative reconstruction error as a function of per-pixel SNR for models trained with the MSE (orange), MAE (blue), and $\chi^2$ (green) reconstruction losses. Solid lines show the per-bin medians, and shaded bands span the 16th--84th percentiles. Lower values are better; the $\chi^2$-trained model attains the lowest median error over most of the range, whereas the MSE-trained model has the largest error in the faint, low-SNR regime.}
\label{fig:snr_error}
\end{figure}

Figure~\ref{fig:snr_error} quantitatively supports the visual differences. At low SNR, the MSE-trained model has the largest median relative error, the MAE-trained model has a smaller error, and the $\chi^2$-trained model performs best over most of the evaluated SNR range. This behavior is consistent with inverse-variance weighting, which reduces the relative influence of high-variance bright pixels and increases that of faint pixels. The improvement of the $\chi^2$ objective therefore results from incorporating pixel-wise uncertainty information rather than from the squared-error form itself. The MSE baseline behaves as expected for an unweighted objective under heteroscedastic noise, where pixels with larger variance can dominate the optimization. The relative errors of all three models exceed $10\%$ at low SNR because weak signals are difficult to recover and division by the small floored denominator $\max(|\mathrm{GT}_i|,\sigma_i)$ amplifies absolute errors. This qualitative ordering is reproduced for two additional training seeds, remains stable across small, medium, and large model capacities, and is further examined at other wavelengths in Appendix~\ref{app:robustness}.

\subsection{Recovery of galaxy structural parameters}
\label{sec:structural_parameters}

To determine whether the pixel-level reconstruction improvement translates into reliable structure measurements, we applied \texttt{statmorph} to the noise-free ground-truth images (GT), the corresponding noisy simulated observations (OBS), and the MAE-, MSE-, and $\chi^2$-VAE reconstructions. For this analysis only, every image was centred and cropped to the same $64\times64$ pixel region. The GT images were processed with the identical centring, cropping, segmentation, and measurement procedure. The Sérsic index was constrained to $0.25\leq n\leq8$, and we retained only paired measurements for which both the GT image and the image being evaluated had \texttt{flag\_sersic}$=0$. This quality cut reduces the sample from 1,211 test-set projections to 1,029 valid GT--OBS pairs and 1,029 valid GT--$\chi^2$-VAE pairs; the corresponding numbers are 1,019 for MAE and 1,006 for MSE. For the half-light radius, Petrosian radius, and Sérsic index, we report the relative error, $100(X_{\rm method}-X_{\rm GT})/X_{\rm GT}$, expressed as a percentage, whereas signed residuals, $X_{\rm method}-X_{\rm GT}$, are used for concentration, asymmetry, smoothness, Gini, and $M_{20}$ because these quantities can approach zero.

\begin{figure*}
    \centering
    \includegraphics[width=\textwidth]{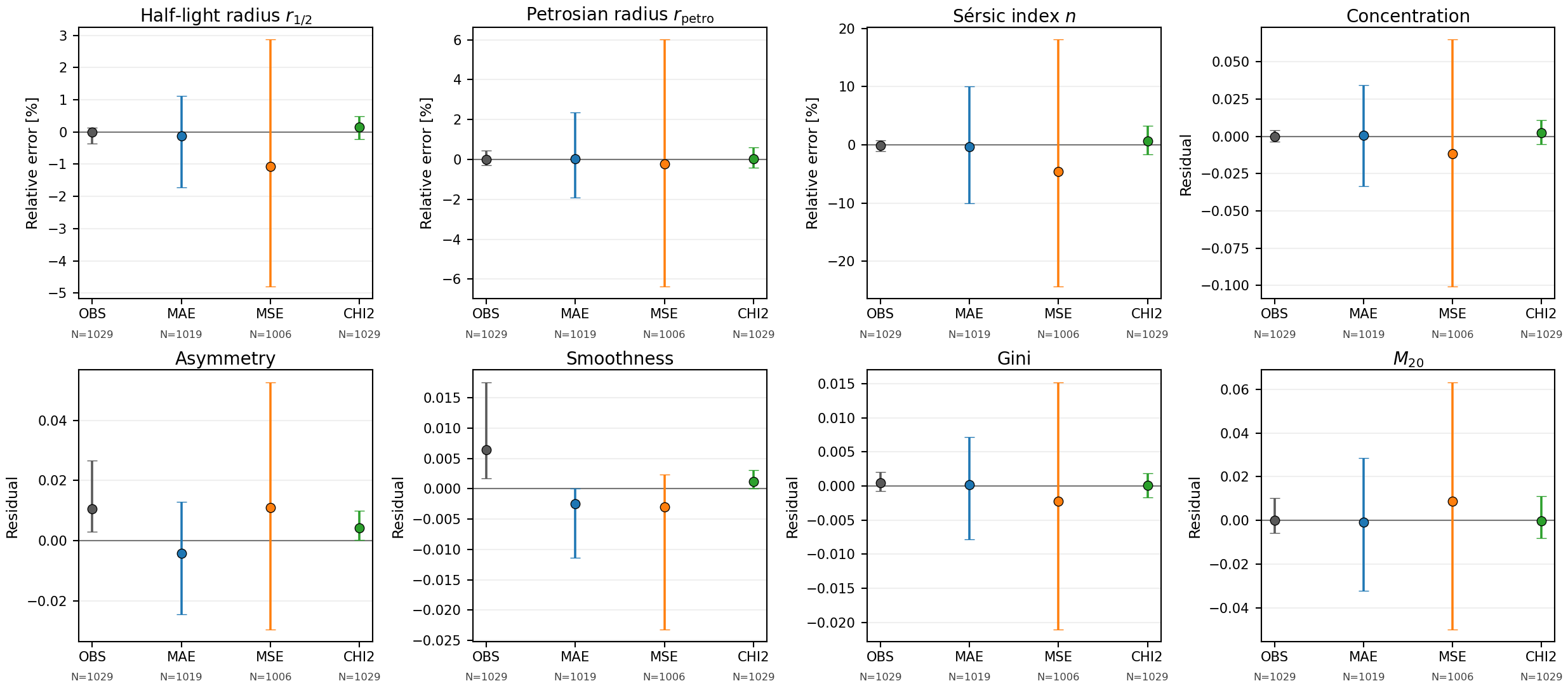}
    \caption{Recovery of structural parameters measured from the centred $64\times64$ pixel test-set stamps. Points show the median residual relative to the noise-free GT measurement, and error bars show the 16th--84th percentile interval. Relative errors are shown for the half-light radius, Petrosian radius, and Sérsic index, while signed residuals are shown for the remaining dimensionless parameters. OBS denotes direct measurement of the noisy simulated observation and therefore provides a non-learning baseline. The number below each point gives the number of valid GT--method pairs after requiring \texttt{flag\_sersic}$=0$ for both images.}
    \label{fig:statmorph_residuals}
\end{figure*}

Figure~\ref{fig:statmorph_residuals} shows that the $\chi^2$-VAE preserves the measured structural properties with small population-level biases and substantially narrower residual distributions than the MAE and MSE reconstructions. Its median relative errors are $0.14\%$, $0.04\%$, and $0.62\%$ for the half-light radius, Petrosian radius, and Sérsic index, respectively. The corresponding 16th--84th percentile intervals are $[-0.23,0.49]\%$, $[-0.40,0.60]\%$, and $[-1.65,3.24]\%$. Concentration, Gini, and $M_{20}$ are likewise recovered with median residuals close to zero. Compared with direct measurements from OBS, the $\chi^2$-VAE reconstructions exhibit modestly broader residual distributions for the radius, Sérsic, concentration, Gini, and $M_{20}$ measurements, while their median residuals remain close to the GT values. Conversely, for asymmetry and smoothness, the $\chi^2$-VAE produces both smaller median offsets and narrower residual distributions than OBS: the median residuals decrease from $0.0105$ and $0.0064$ for OBS to $0.0042$ and $0.0012$ for the $\chi^2$-VAE, respectively. Thus, the overall structural-parameter recovery of the $\chi^2$-VAE is broadly comparable to that obtained directly from OBS, although their relative performance depends on the parameter considered. In contrast, the MAE and particularly the MSE reconstructions exhibit much broader distributions, including a Sérsic-index interval of approximately $[-24.3,18.1]\%$ for MSE. These results show that the pixel-level advantage of the $\chi^2$ objective over MAE and MSE is also reflected in more stable structural measurements, while no substantial population-level morphological offsets are introduced within the parameters examined here.

\subsection{Real DESI Observations}
\label{sec:desi_results}

\begin{figure}[t]
\centering
\includegraphics[width=\linewidth]{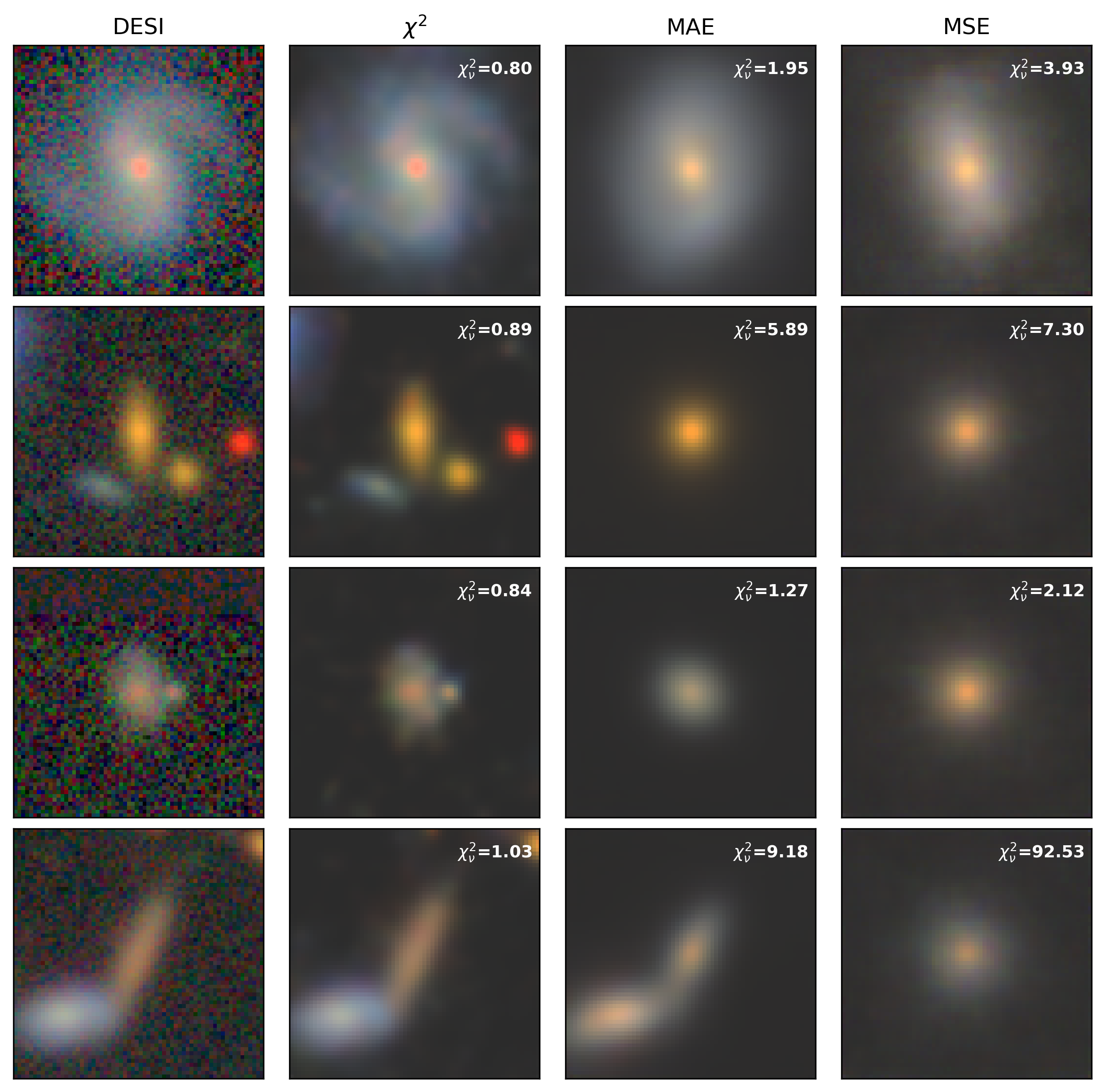}
\caption{Reconstructions of four randomly selected real observations from the DESI Bright Galaxy Survey test set, shown as the observation (DESI) and the models trained from scratch with the $\chi^2$, MAE, and MSE reconstruction losses on the $g$, $r$, and $z$ images. Each reconstruction is annotated with its reduced $\chi^2_\nu$ against the observation.}
\label{fig:desi}
\end{figure}
We demonstrate the method qualitatively on real observations by training models from scratch with the MSE, MAE, and $\chi^2$ reconstruction losses on the Legacy Surveys $g$, $r$, and $z$ images of galaxies in the Dark Energy Spectroscopic Instrument (DESI) Bright Galaxy Survey and applying them to the same test set (Figure~\ref{fig:desi})\cite{dey_overview_2019a}. The model architecture, capacity, and other hyperparameters are kept identical to those adopted in Section~\ref{sec:experiments}; because the real DESI training sample is substantially larger than the simulated sample, the total number of training steps is increased from 60,000 to 300,000 to allow the models to converge on the larger dataset. Because these images have no noise-free GT, they do not permit the quantitative comparison performed on the simulations.

To evaluate the three models without a GT, we compute, for each three-band image, the reduced $\chi^2$ diagnostic
\begin{equation}
\chi^2_\nu=\frac{1}{N_{\mathrm{valid}}}
\sum_{i=1}^{N_{\mathrm{valid}}}\frac{(O_i-P_i)^2}{\sigma_i^2},
\qquad N_{\mathrm{valid}}=64\times64\times3,
\end{equation}
where $\sigma_i$ is derived from the survey inverse-variance map and the number of degrees of freedom is taken to equal the number of valid pixels. A value close to unity indicates that the reconstruction is statistically consistent with the observation given the noise model.

Using this diagnostic, the three objectives show clear qualitative differences on the same four randomly selected test galaxies. The MAE-trained reconstruction reproduces the broad light profile of the four systems but lacks fine detail, and its colors show noticeable deviations from the observation. The MSE-trained reconstruction performs worse still: its $\chi^2_\nu$ values are larger than those of the MAE model for all four examples. In comparison, the $\chi^2$-trained model achieves $\chi^2_\nu$ values closest to unity, namely 0.80, 0.89, 0.84, and 1.03 for the four examples. Among these four selected examples, the first reconstruction retains a compact central component and diffuse outer emission; the second recovers several nearby sources with distinct colors; the third recovers a faint compact source adjacent to the diffuse main object; and the fourth retains both components and the elongated structure of the system.

\section{Discussion}
\label{sec:discussion}
Our quantitative conclusions apply to a controlled observation model with a fixed Moffat PSF (FWHM $1.5''$), a fixed pixel scale, a weakly tilted sky background, Poisson--Gaussian detector noise, and mesh-based background estimation. Because the GT is PSF-convolved, the comparison assesses signal recovery at the adopted instrumental resolution. The Gaussian likelihood is approximate at very low counts, and detector defects, cosmic rays, and resampling-induced noise correlations are not modeled. Real PSFs can depart from a single analytic profile through asymmetry, extended wings, and spatial or temporal variations, redistributing light and altering feature contrast and structural measurements. The separately trained models on real DESI Legacy Surveys images (Section~\ref{sec:desi_results}) show similar qualitative differences in reconstruction quality, providing some support for applicability to observed data. These examples do not, however, isolate robustness to PSF variation or establish quantitative signal-recovery accuracy without a noise-free GT. Space-based images can have substantially narrower PSFs than adopted here, potentially resolving finer structures and changing the pixel-wise signal and noise distributions. Their reconstruction quality and the relative performance of the losses also depend on sampling and instrument-specific PSF structure. Quantifying these effects requires dedicated space-based PSF, sampling, and noise models and is beyond the scope of this study.

We use a VAE because its latent-variable formulation supports both image reconstruction and generation while learning a compressed representation of galaxy morphology. We did not test alternative network architectures, so whether the relative behavior of the three reconstruction losses generalizes beyond the VAE remains unknown. The number of distinct simulated galaxies is limited by the cost of the SKIRT calculations, each of which uses approximately $10^{8}$ photon packets to reduce Monte Carlo radiative-transfer noise; random cropping increases the number of training examples but not the number of independent galaxy morphologies. Although Section~\ref{sec:structural_parameters} shows that the pixel-level advantage is also reflected in more stable structural measurements, this assessment is limited to the parametric and non-parametric morphological indicators examined there.

The comparison also depends on the training configuration. Although the models share the same architecture and optimization settings, the different numerical scales of the reconstruction losses imply that a common KL coefficient does not provide matched effective regularization. Checkpoints are also selected using each model's own validation reconstruction metric. In the simulated $r$-band experiment, reducing the MAE and MSE KL weights leaves the relative performance ordering unchanged, but does not equalize the realized weighted-KL-to-reconstruction ratios (Appendix~\ref{app:robustness}). This sensitivity test supports the result under the configurations examined, while leaving open whether individually optimized models would show the same differences. Establishing the optimal configuration for each objective, including training length and KL weighting, would require a dedicated hyperparameter search beyond the scope of this comparison.

\section{Conclusion}
\label{sec:conclusion}
We compared the MSE, MAE, and $\chi^2$ reconstruction losses in otherwise identical VAEs and evaluated their outputs against SKIRT-based GT images. The $\chi^2$-trained model has the lowest median relative error over most of the evaluated SNR range, whereas the MSE-trained model has the largest error in faint pixels. Within the simulated observation process and VAE architecture considered here, these results indicate that inverse-variance weighting improves the recovery of astronomical signals under heteroscedastic, low-SNR conditions. The four DESI examples further show qualitatively that a separately trained $\chi^2$ model reproduces visible multicomponent and diffuse structures while leaving predominantly noise-like residuals.

%%%%%%%%%%%%%%%%%%%%%%%%%%%%%%%%%%%%%%%%%%
\vspace{6pt}

%%%%%%%%%%%%%%%%%%%%%%%%%%%%%%%%%%%%%%%%%%
\authorcontributions{Conceptualization, R.Y.; methodology, R.Y.; software, R.Y.; validation, R.Y.; formal analysis, R.Y., Q.X. and S.S.; investigation, R.Y.; data curation, R.Y.; writing---original draft preparation, R.Y.; writing---review and editing, Q.X. and S.S.; visualization, R.Y. All authors have read and agreed to the published version of the manuscript.}

\funding{This work was supported by the National Key R\&D Program of China
(Nos.\ 2022YFF0503402 and 2019YFA0405501), the National Natural Science
Foundation of China (Nos.\ 12073059 and 12141302), the Shanghai
Academic/Technology Research Leader Program (22XD1404200), and the
China Manned Space Project (CMS-CSST-2021-A07).}

\institutionalreview{Not applicable.}

\informedconsent{Not applicable.}

\dataavailability{The code and data supporting the results of this study are available at \url{https://github.com/Rh-YE/obs2gt}.}

\acknowledgments{The DESI Legacy Imaging Surveys consist of three complementary projects: the Dark Energy Camera Legacy Survey (DECaLS), Beijing--Arizona Sky Survey (BASS), and Mayall $z$-band Legacy Survey (MzLS). DECaLS, BASS, and MzLS include data obtained, respectively, with the Blanco telescope at the Cerro Tololo Inter-American Observatory (CTIO), operated by NSF's National Optical-Infrared Astronomy Research Laboratory (NOIRLab); the Bok telescope at Steward Observatory; and the Mayall telescope at Kitt Peak National Observatory (KPNO), operated by NOIRLab. Pipeline processing was supported by NOIRLab and Lawrence Berkeley National Laboratory (LBNL). This project used data obtained with the Dark Energy Camera (DECam), constructed by the Dark Energy Survey (DES) collaboration; funding for DES was provided by the U.S.\ Department of Energy (DOE), the National Science Foundation (NSF), and numerous international agencies (full list at \url{https://www.darkenergysurvey.org}). BASS is a key project of the Telescope Access Program (TAP), funded by the National Astronomical Observatories, Chinese Academy of Sciences (NAOC), and the Chinese Academy of Sciences. The Legacy Surveys team thanks the Tohono O'odham Nation for access to Iolkam Du'ag (Kitt Peak). NOIRLab is operated by the Association of Universities for Research in Astronomy (AURA) under a cooperative agreement with NSF; LBNL is managed by the Regents of the University of California under contract to the U.S.\ Department of Energy. The Legacy Surveys imaging of the DESI footprint is supported by DOE Contract No.\ DE-AC02-05CH1123 and NSF Contract No.\ AST-0950945.}

\conflictsofinterest{The authors declare no conflicts of interest.}

%%%%%%%%%%%%%%%%%%%%%%%%%%%%%%%%%%%%%%%%%%
%% Optional

%% Only for journal Encyclopedia
%\entrylink{The Link to this entry published on the encyclopedia platform.}

\abbreviations{Abbreviations}{%
The following abbreviations are used in this manuscript:\\

\noindent
\begin{tabular}{@{}ll}
CSST & China Space Station Telescope\\
DESI & Dark Energy Spectroscopic Instrument\\
GT & Ground truth\\
KL & Kullback--Leibler\\
MAE & Mean absolute error\\
MSE & Mean squared error\\
PSF & Point-spread function\\
RMS & Root mean square\\
RN & Read noise\\
SNR & Signal-to-noise ratio\\
VAE & Variational autoencoder\\
ZP & Photometric zero point
\end{tabular}
}

\appendix
\section{Implementation Details}
\label{app:implementation}
The SKIRT source images are rendered at $100~\mathrm{pc\,pixel^{-1}}$ before the forward model is applied.

For the simulated $r$-band comparison, the reconstruction network is a single-channel convolutional VAE operating on $64\times64$-pixel input and output images, with a diagonal-Gaussian latent regularizer. It uses two residual blocks per resolution level and channel multipliers $[1,2,4]$, with a base channel width of 128 and an $80\times16\times16$ latent representation, giving 54.3 million parameters. The decoder uses a softplus output activation. For the $\chi^2$ model, the uncertainty map is supplied only to the reconstruction loss and is not concatenated with the input; the encoder receives only the noisy observation.

All three simulated-data variants are trained for 60,000 optimization steps with the Adam optimizer\cite{kingma_adam_2015} at a fixed learning rate of $10^{-4}$ and a training batch size of 64. Validation is performed every 250 training steps with a batch size of 32 and is limited to 100 batches. Image values remain in units of nanomaggies without min--max normalization or standardization.
\begin{figure}[t]
\centering
\includegraphics[width=\textwidth]{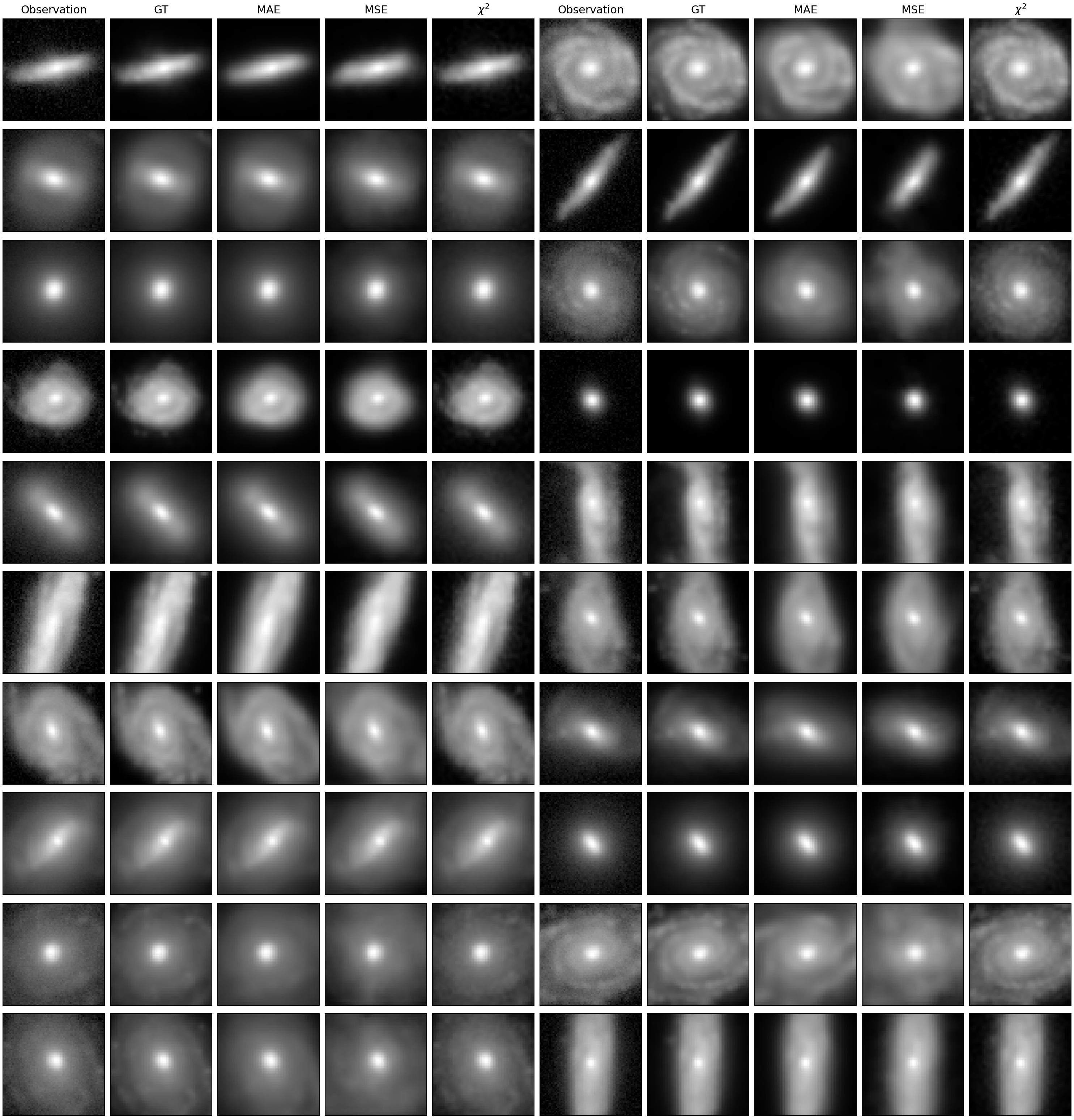}
\caption{More randomly selected test galaxies. Columns show the noisy observation, GT, and reconstructions produced by models trained with the MAE, MSE, and $\chi^2$ losses. The same color scale is used within each row. For display, the transformation $\operatorname{arcsinh}(50I)/\sqrt{50}$ is applied to each image $I$.}
\label{fig:more}
\end{figure}
\section{More examples}
We show more randomly selected test galaxies in Fig.\ref{fig:more}.

% \section{Robustness to Model Capacity, Wavelength, and KL Weight}
\label{app:robustness}

We performed additional experiments to determine whether the relative performance of the three reconstruction objectives depends on the model capacity, observed wavelength, or effective strength of the KL regularization. Unless stated otherwise, all experiments use the same per-pixel relative-error definition, data selection, SNR binning, and checkpoint-selection procedure as the primary $r$-band comparison in Figure~\ref{fig:snr_error}.

\subsection{Model Capacity}

\begin{figure*}[t]
\centering
\includegraphics[width=\textwidth]{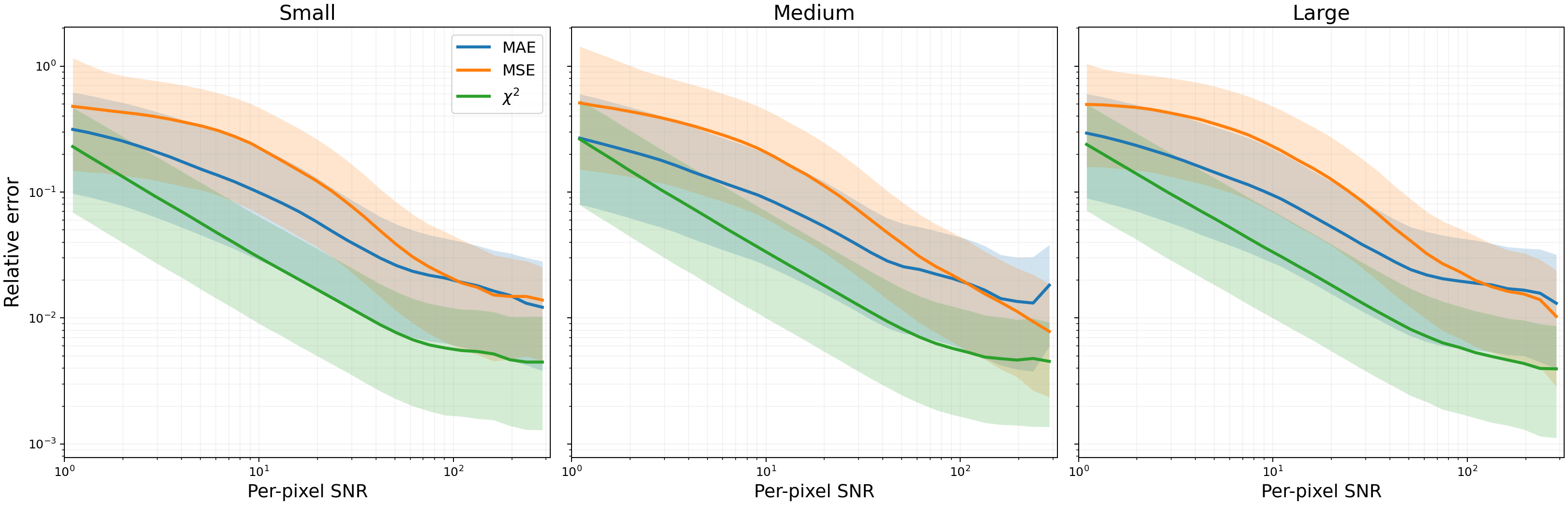}
\caption{Same as Figure~\ref{fig:snr_error}, but for the small, medium, and large model configurations trained with random seed 42.}
\label{fig:appendix_capacity}
\end{figure*}

We repeated the comparison using small, medium, and large model configurations at fixed random seed 42. As shown in Figure~\ref{fig:appendix_capacity}, changing the model capacity modifies the precise relative-error curves but does not change their qualitative ordering. The $\chi^2$-trained model achieves the lowest median relative error over most of the evaluated SNR range for all three capacities. This indicates that its reconstruction advantage is not specific to the adopted 54.3-million-parameter architecture.

\subsection{Observed Wavelength}

\begin{figure*}[t]
\centering
\includegraphics[width=\textwidth]{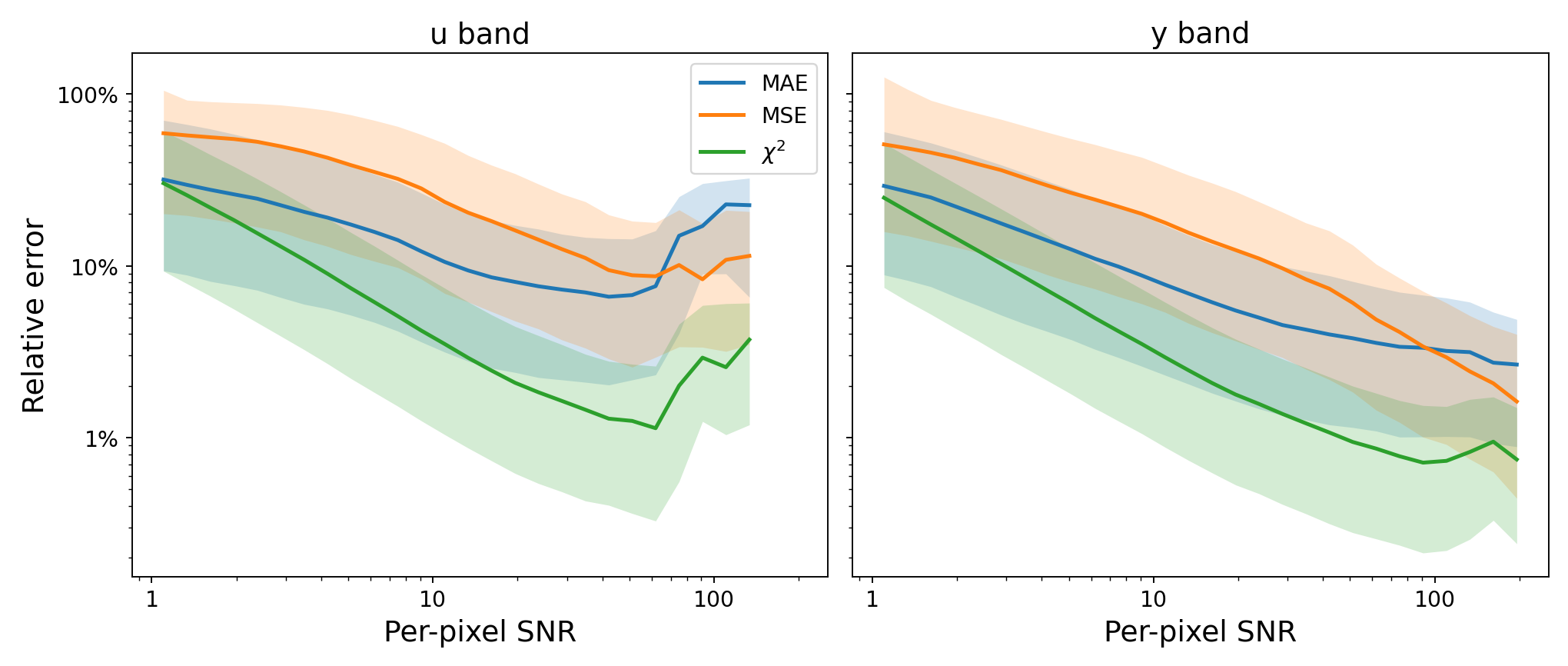}
\caption{Same as Figure~\ref{fig:snr_error}, but for the CSST $u$- and $y$-band models trained with random seed 42.}
\label{fig:appendix_band}
\end{figure*}

We further repeated the comparison in two additional CSST bands, $u$ and $y$, using the large model configuration and random seed 42. Figure~\ref{fig:appendix_band} shows that the qualitative ordering is preserved at both wavelengths: the $\chi^2$-trained model achieves the lowest median relative error over most of the SNR range, followed by the MAE- and MSE-trained models. The advantage of inverse-variance weighting is therefore not restricted to the $r$ band and persists for galaxy images with different wavelength-dependent flux distributions.

\subsection{KL-Weight Sensitivity}
\label{app:kl_sensitivity}

Because the MAE, MSE, and $\chi^2$ reconstruction terms have different numerical scales, using the same KL coefficient does not necessarily produce the same effective balance between reconstruction fidelity and latent regularization. We evaluated the realized loss terms at the selected checkpoints using the same 4,844 $r$-band validation crops. For the primary models with $\beta=10^{-4}$, the mean reconstruction terms were $0.0314$, $0.00790$, and $0.910$ for MAE, MSE, and $\chi^2$, respectively, while the corresponding weighted KL terms, $\beta D_{\mathrm{KL}}$, were $0.0133$, $0.00955$, and $0.115$. The resulting weighted-KL-to-reconstruction ratios were $0.42$, $1.21$, and $0.13$, confirming that a common value of $\beta$ does not provide matched effective regularization across the three objectives.

\begin{figure}[t]
\centering
\includegraphics[width=0.9\linewidth]{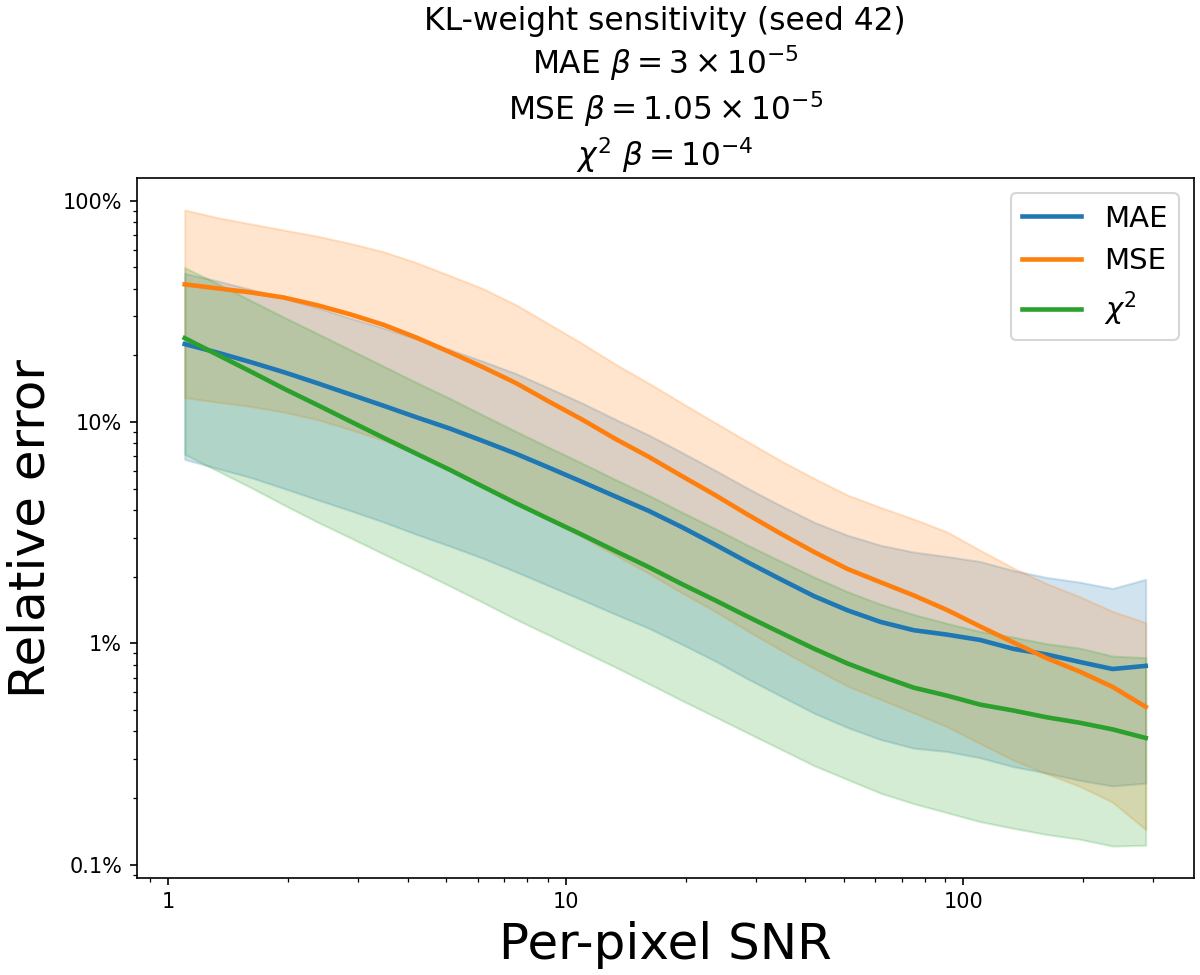}
\caption{Same as Figure~\ref{fig:snr_error}, but using the MAE and MSE models retrained with reduced KL weights of $\beta=3\times10^{-5}$ and $\beta=1.05\times10^{-5}$, respectively. The $\chi^2$ model retains the primary value of $\beta=10^{-4}$.}
\label{fig:appendix_kl}
\end{figure}

As a sensitivity test, we retrained the MAE and MSE models from scratch using reduced KL weights of $\beta=3\times10^{-5}$ and $\beta=1.05\times10^{-5}$, respectively, while retaining $\beta=10^{-4}$ for the $\chi^2$ model. The realized weighted-KL-to-reconstruction ratios became $0.32$ for MAE and $1.22$ for MSE. Thus, directly rescaling $\beta$ does not enforce a matched ratio because the reconstruction and latent terms jointly readjust during optimization. Nevertheless, Figure~\ref{fig:appendix_kl} shows that the relative performance ordering remains unchanged, with the $\chi^2$ model retaining the lowest median relative error over most of the evaluated SNR range. The main result is therefore insensitive to this reduction of the MAE and MSE KL weights, although identifying the individually optimal rate--distortion balance for each objective would require a dedicated hyperparameter search.

%%%%%%%%%%%%%%%%%%%%%%%%%%%%%%%%%%%%%%%%%%
%\isPreprints{}{% This command is only used for ``preprints''.
\begin{adjustwidth}{-\extralength}{0cm}

\reftitle{References}

\bibliography{example_paper}

\PublishersNote{}
%\isPreprints{}{% This command is only used for ``preprints''.
\end{adjustwidth}

\end{document}